\documentstyle[11pt,newpasp,twoside,epsfig]{article}
\begin{document}

\title{XMM-Newton Observations of AGN Iron Line profiles}
\author{James Reeves}
\affil{University of Leicester, UK}

\setcounter{page}{1}
\index{Reeves, J.}

\begin{abstract}

XMM-Newton observations of type I AGN are presented. 
The properties of the iron K emission line are reviewed,
the majority of AGN observed by XMM-Newton show narrow, unresolved iron 
lines at 6.4 keV from cold matter 
that must originate far from the inner accretion disc, perhaps in the 
putative torus or outer broad line region. 
The strength of this narrow line appears to decrease with luminosity, 
implying a reduction in the covering fraction of this material in the more
luminous quasars. Few examples of the broad, relativistic iron line
profile have been found by XMM-Newton, although in MCG -6-30-15, the extreme
breadth of the broad line component may imply a 
Kerr metric for the central black hole. Generally,
relativistic Fe K line profiles are not required in a number of other
Seyfert~1 X-ray spectra.  

\end{abstract}

\vspace{-1.0cm}
\section{Introduction}
Since the iron K$\alpha$ emission line was first found to be common in
the X-ray spectra of AGN, it has been recognized as an important probe
of the matter nearest to the central engine. 
Hard X-ray observations with {\it Ginga} 
first established that iron K$\alpha$ emission was predominant in
Seyfert 1 galaxies, and often accompanied by a `cold' iron K edge at
$\sim$7~keV and a Compton scattered `hump' at higher energies (Pounds
et al. 1990, Nandra \& Pounds 1994). A favoured interpretation was
that hard X-rays were being `reflected' by Compton thick matter, in
the form of an accretion disc subtending a solid angle of $\sim2\pi$
to the primary X-ray source. 

The {\it ASCA} X-ray observatory subsequently resolved the iron
K$\alpha$ emission in one Seyfert 1, MCG~-6-30-16 (Tanaka et
al. 1995). The line profile appeared to be extremely broad 
(with $v\sim0.3c$) and
subject to distortions from relativistic
effects a few gravitational radii away from a massive black
hole. Subsequent {\it ASCA} observations appeared to show that the broad
iron K line profile was common amongst other Seyfert 1 galaxies
(Nandra et al. 1997), although in a recent re-analysis
of archival {\it ASCA} data, Lubinski \& Zdziarski (2001) have claimed that
the mean Seyfert 1 iron line profile may be narrower than previously
believed. 

The successful launch of XMM-Newton will now help to revolutionize the 
study of Active Galactic Nuclei in the X-ray waveband. 
In particular, the considerable collecting area of the XMM-Newton EPIC
detectors at 6 keV, when
compared to {\it ASCA} or even {\it Chandra}, makes XMM-Newton 
{\it the} X-ray observatory with which to study the iron K line profile.

\begin{figure*} 
\centering
\vbox{
\includegraphics[width=5cm, height=6.5cm, angle=270]{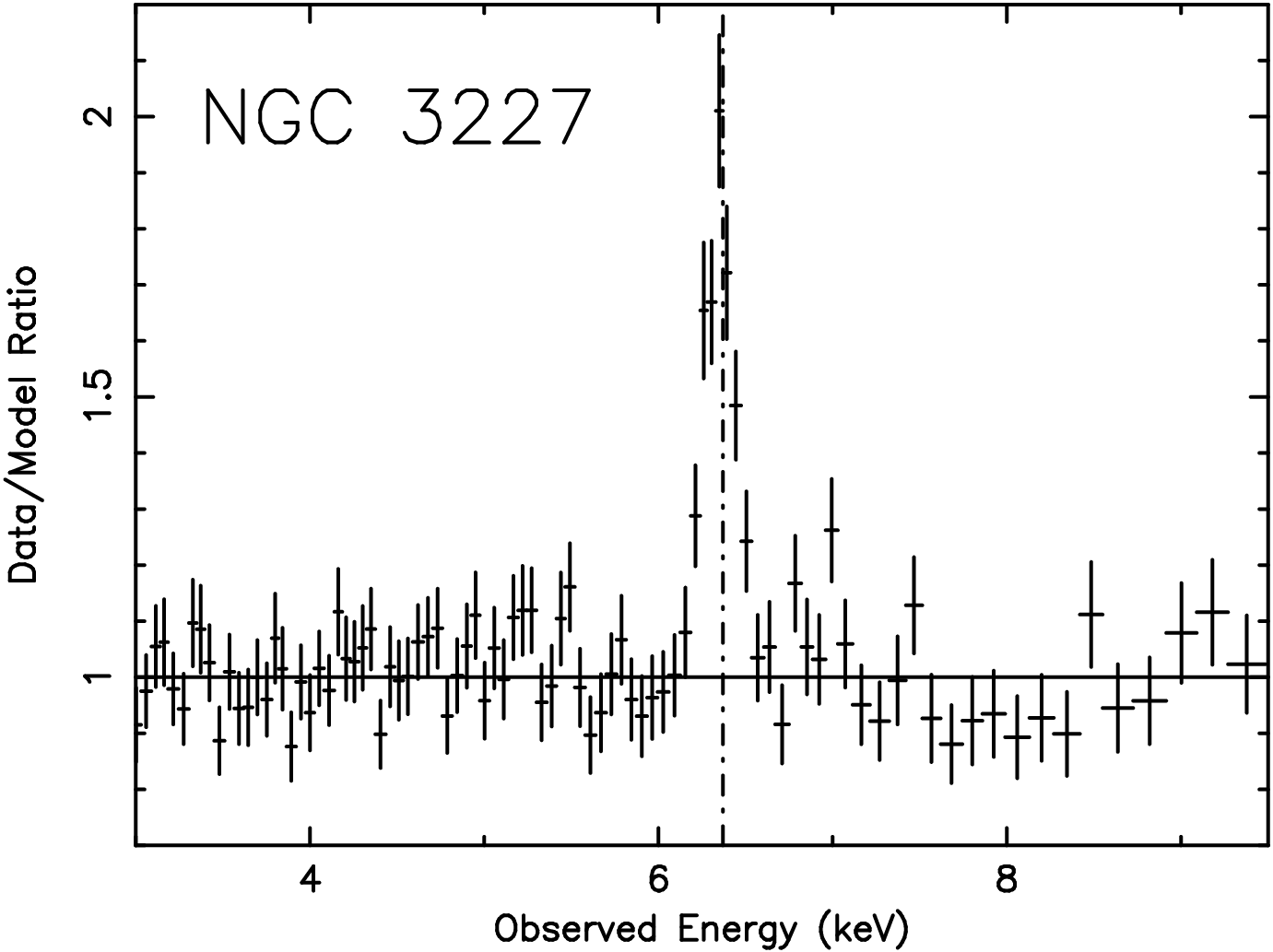}
\includegraphics[width=5cm, height=6.5cm, angle=270]{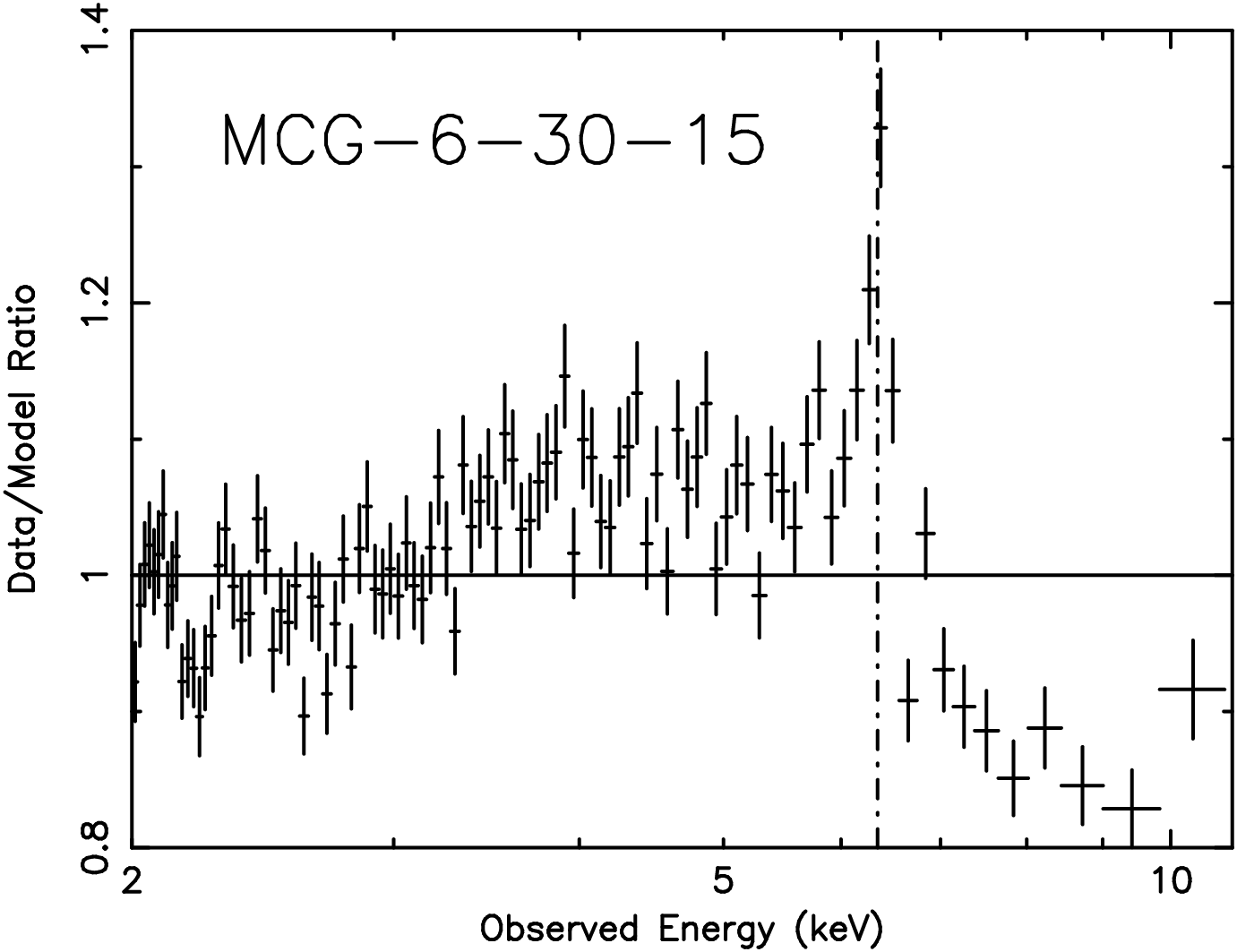}
\includegraphics[width=5cm, height=6.5cm, angle=270]{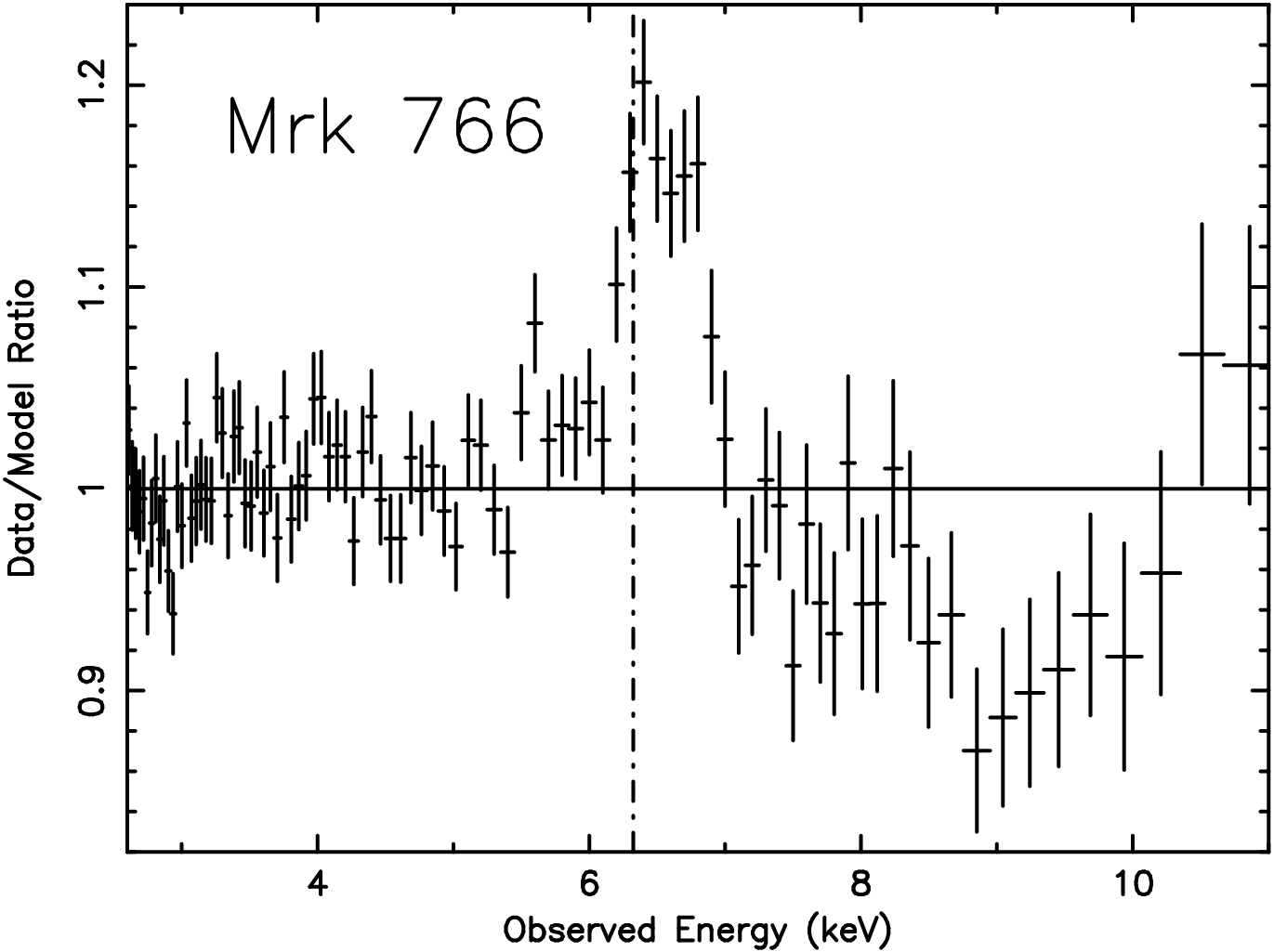}
\includegraphics[width=5cm, height=6.5cm, angle=270]{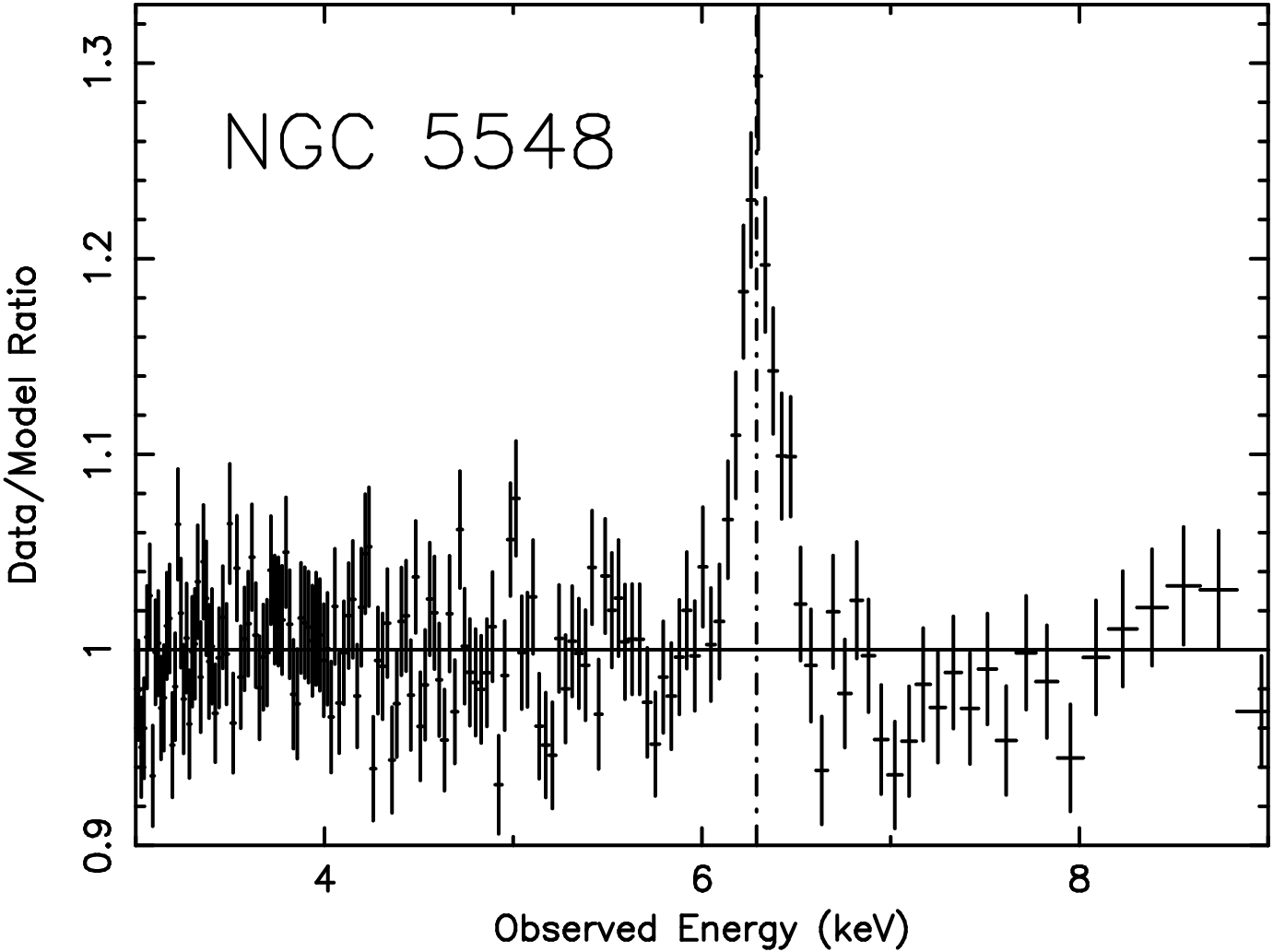}
\includegraphics[width=5cm, height=6.5cm, angle=270]{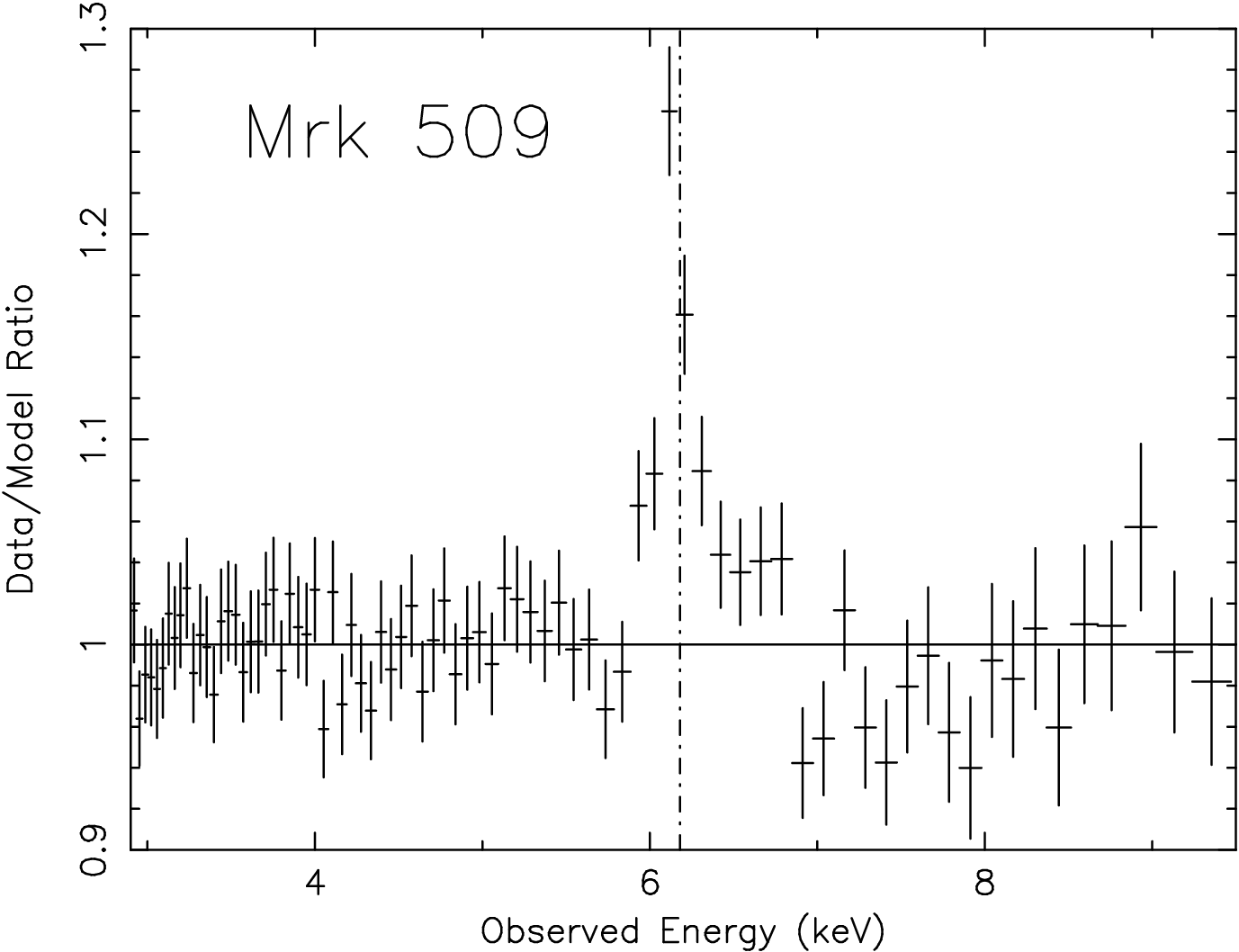}
\includegraphics[width=5cm, height=6.5cm, angle=270]{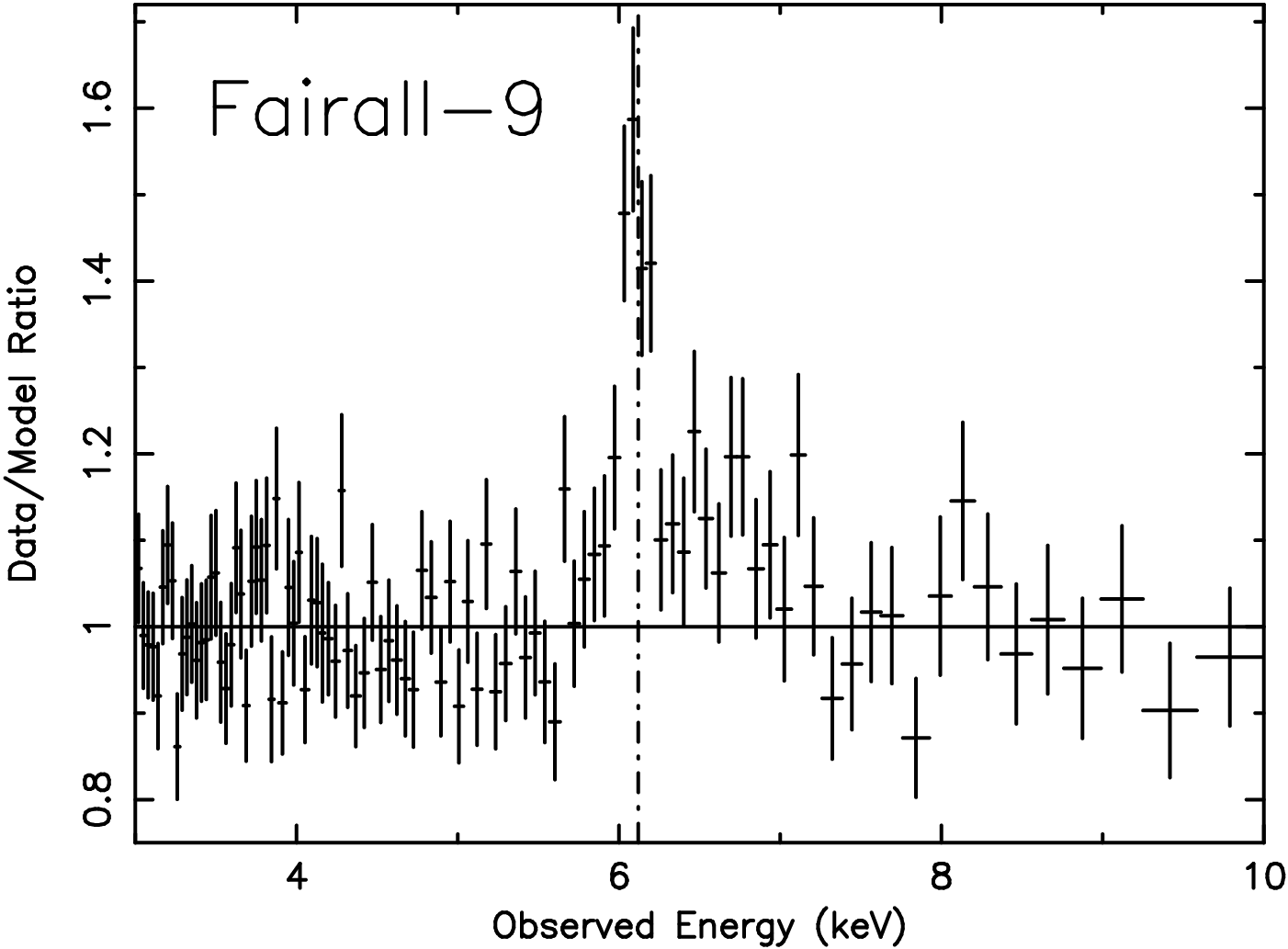}
\includegraphics[width=5cm, height=6.5cm, angle=270]{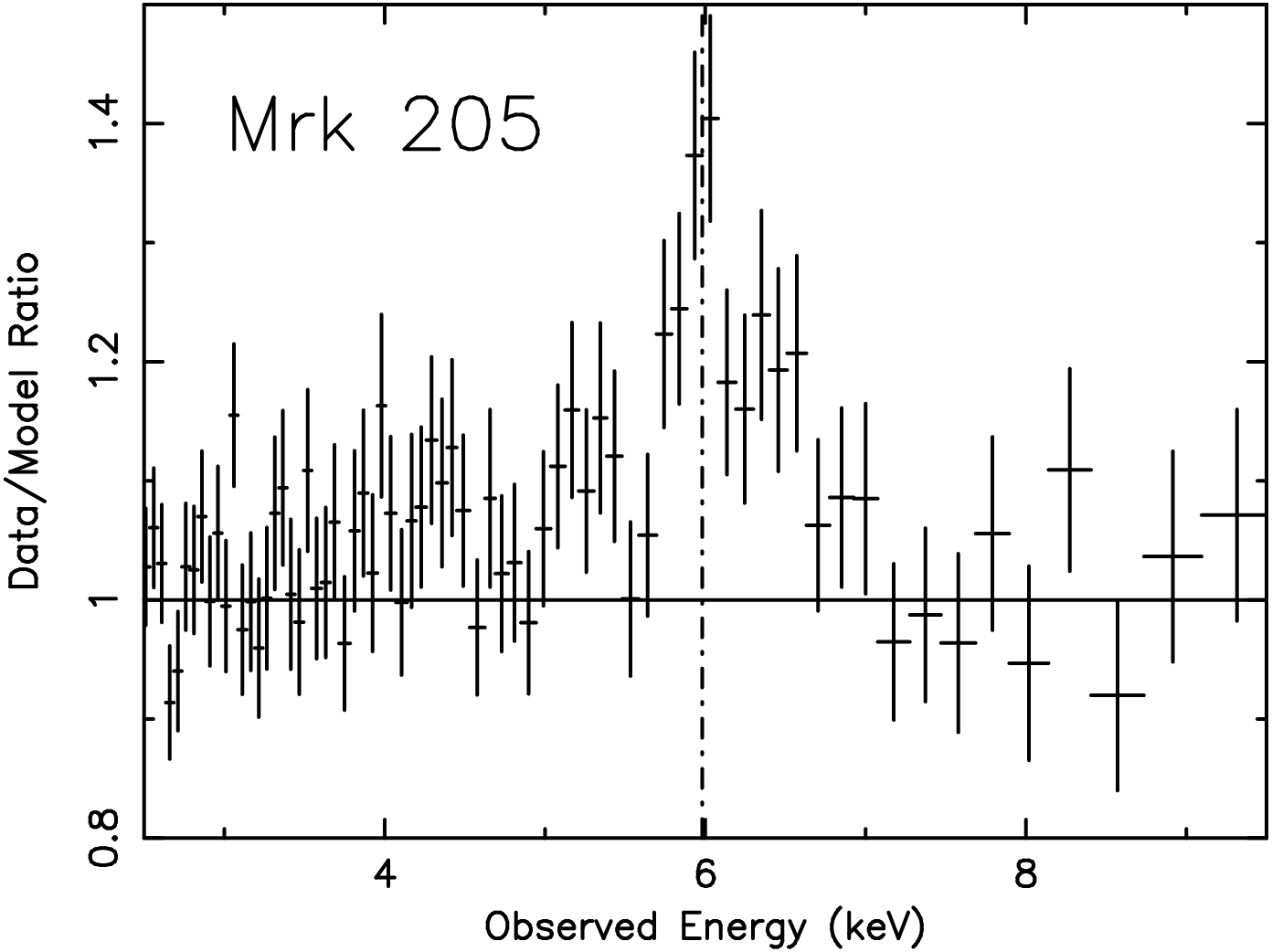}
\includegraphics[width=5cm, height=6.5cm, angle=270]{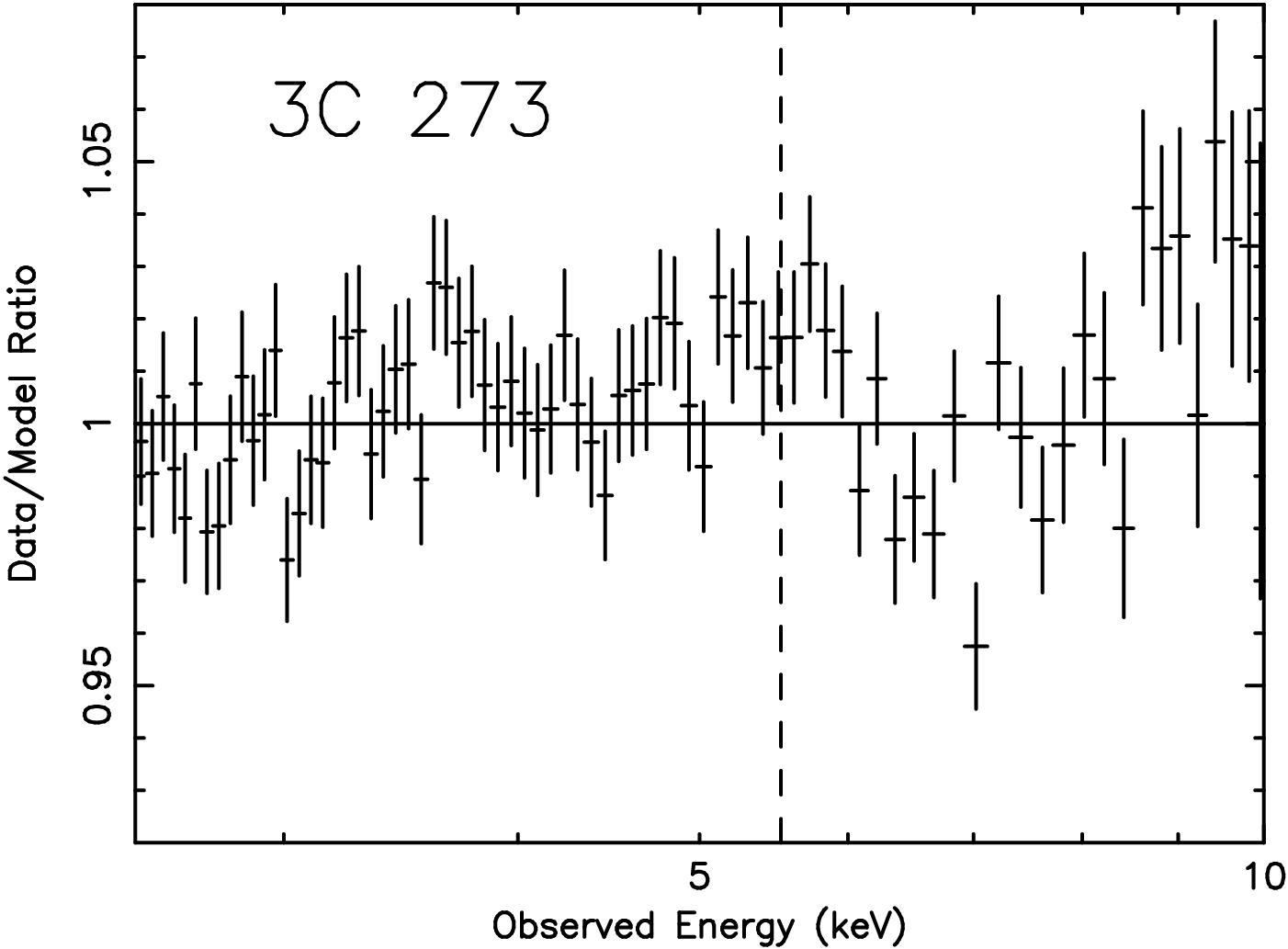}
}
\caption{Iron line profiles of 8 Type 1 AGN, measured by XMM-Newton
EPIC-PN. The vertical dot-dashed line shows the position of the
neutral 6.4 keV Fe line energy, in the object rest frame.}
\end{figure*}

\section{The Iron line Profile of Type I AGN with XMM-Newton}

This paper presents early observations of type I AGN from XMM-Newton;
here the iron K emission line profiles of a 
subsample of 8 type I AGN are examined.
The X-ray spectra were fitted between 2.5 and 12 keV, in
order to determine the hard X-ray continuum. 
The iron line profiles are plotted in Figure 1, showing the
data/model ratio to an absorbed power-law fit, ordered by
increasing 2-10 keV luminosity.
The narrow iron line component at 6.4 keV appears common in all but one
of the AGN (the exception being in the most luminous AGN, 3C~273). The
mean energy of the narrow iron line component is $6.39\pm0.02$~keV,
indicating a low ionisation ($<$Fe~\textsc{XV}). 
Generally the line cores are 
unresolved in the EPIC CCD spectra, the FWHM resolution at 6.4 keV is
typically $<120$~eV, corresponding to 
$<5000$~km~s$^{-1}$ for the (FWHM) line velocity width. 

The mean equivalent
width of the narrow component is
$\sim75$~eV. Typically the datasets also require the addition of a cold
reflection component, with covering fraction
$R=\Omega/2\pi\sim0.5$. The relative strengths of the
narrow iron line and reflection components are consistent with 
reflection off distant Compton-thick matter, subtending a solid angle of
$\sim\pi$ steradian to the X-ray source (Reeves et al. 2001). 
Within the framework of AGN
unification schemes, one possible location for this
material is the molecular torus. Interestingly the 
Chandra-HETGS line profile of the Seyfert~1 NGC 3783 (Kaspi et al. 2002) 
is resolved with a FWHM velocity of
$\sim1700$~km~s$^{-1}$, consistent with an origin from the 
torus. On the other hand, a Chandra grating observation of NGC 5548
(Yaqoob et al. 2001) measured a FWHM width of
$\sim4500$~km~s$^{-1}$, 
indicating a contribution from the broad line region. 
Future observations at higher resolution
with calorimeter-based detectors 
(for example the XRS on ASTRO-E2), 
with resolution $\Delta E/E\sim1000$ at 6 keV, are required in
order to resolve the various components of the narrow iron line core.

\begin{figure*} 
\centering
\vbox{
\includegraphics[width=6.5cm, height=10.5cm, angle=270]{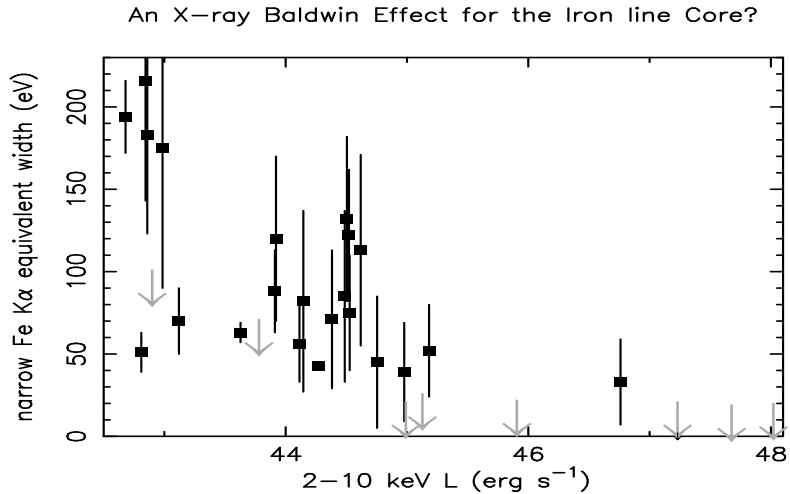}
}
\caption{Narrow iron K line equivalent width versus luminosity, 
for type I AGN measured with XMM-Newton. 
An X-ray Baldwin effect is observed, where the
line strength falls with increasing luminosity.}
\end{figure*}

It is interesting to note that the equivalent width of the
narrow iron line appears to diminish with luminosity - the so-called
X-ray Baldwin effect. 
This is illustrated in Figure 2 (Kim Page, private comm), which plots the 
iron line core equivalent width against 2-10 keV luminosity for 
various type I AGN observed by XMM-Newton. 
This trend would appear to require a reduction 
in the solid angle of the Compton-thick matter (such as the
torus), responsible for the narrow iron line emission. 
One implication of this is a possible 
reduction in the number density of luminous, Compton-thick 
type II quasars compared with Seyfert 2s. 

Finally a second broad Gaussian emission line component was added 
to the datasets, to constrain any possible relativistic iron K line
component 
associated with emission from the inner accretion disc. The only line
profile (Figure 1) where a significant red-wing (down to 3 keV) 
was found was in MCG~-6-30-15, where the extreme red-shift 
requires a Kerr metric
for the line profile (Wilms et al. 2001). However,
given the breadth of the line, care has to be taken on where the hard
X-ray continuum is fitted in relation to the X-ray warm absorber. 
Simultaneous high energy measurements (Beppo-SAX or Integral)
along with XMM-Newton data can help to resolve this issue (e.g. Fabian et
al. 2002).  

However, relativistic profiles are not formally required in the other
7 objects. Nonetheless, significant ionised iron emission above 
$>6.4$~keV may be present (in Mrk 766, Mrk 205 and Fairall-9), 
indicating that an
ionised disc reflection component may be important (Reeves et
al. 2001; Pounds et al. 2002, in prep). 
In the other objects (NGC 3227, NGC 5548, Mrk
509, 3C 273) an additional broad line component is not
required, e.g. the formal $3\sigma$ upper limit on the equivalent width of
a relativistic line component in NGC 5548 is $<43$~eV (Pounds
et al. 2002). 

\vspace{-0.5cm}
\section{Conclusions} 
The iron line profiles of 8 type I AGN were examined with
XMM-Newton. Its was found that a narrow line core appears ubiquitous in
the Seyfert~1s, the exception being in the 
highest luminosity AGN. However the relativistic, redshifted profile
associated with emission from the inner disc appears 
less common, here the only apparent case is
MCG~-6-30-15. A systematic study of a large sample of 
Seyfert~1 galaxies with XMM-Newton will be crucial for determining
the relative importance of the relativistic iron line component.

\end{document}